\documentstyle[12pt,aaspp4,epsfig]{article}
\newcommand{\dsp}{\displaystyle}
\def\ref#1{\hangindent=6em\hangafter=1 {#1}}
\def\spose#1{\hbox to 0pt{#1\hss}}
\def\lta{\mathrel{\spose{\lower 3pt\hbox{$\mathchar"218$}}
     \raise 2.0pt\hbox{$\mathchar"13C$}}}
\def\gta{\mathrel{\spose{\lower 3pt\hbox{$\mathchar"218$}}
     \raise 2.0pt\hbox{$\mathchar"13E$}}}

\begin{document}

\title{Magnetic Fields in Massive Stars.\\ II. The Buoyant Rise of Magnetic Flux
Tubes\\ Through the Radiative Interior}

\author{K. B. MacGregor \altaffilmark{1} and J. P.
Cassinelli \altaffilmark{1,2}}

\altaffiltext{1}{High Altitude Observatory, National Center for 
Atmospheric Research, P.O. Box 3000, Boulder, CO 80307-3000;
kmac@hao.ucar.edu}
\altaffiltext{2}{Department of Astronomy, University of Wisconsin, 
475 N. Charter St., Madison WI 53706-1582; cassinelli@astro.wisc.edu}

\begin{abstract} 
We present results from an investigation of the dynamical behavior
of buoyant magnetic flux rings in the radiative interior of a uniformly
rotating early-type star.  Our physical model describes a thin, axisymmetric,
toroidal flux tube that is released from the outer boundary of the convective
core, and is acted upon by buoyant, centrifugal, Coriolis, magnetic tension,
and aerodynamic drag forces.  We find that rings emitted in the equatorial plane
can attain a stationary equilibrium state that is stable with respect to small 
displacements in radius, but is unstable when perturbed in the meridional direction. 
Rings emitted at other latitudes travel toward the surface along trajectories that
largely parallel the rotation axis of the star.  Over much of the ascent, the 
instantaneous rise speed is determined by the rate of heating by the absorption of
radiation that diffuses into the tube from the external medium.  Since the time
scale for this heating varies like the square of the tube cross-sectional radius, 
for the same field strength, thin rings rise more rapidly than do thick rings.
For a reasonable range of assumed ring sizes and field strengths, our results 
suggest that buoyancy is a viable mechanism for bringing magnetic flux from the
core to the surface, being capable of accomplishing this transport in a time that
is generally much less than the stellar main sequence lifetime.

\end{abstract}


\section{Introduction}
Although magnetic fields and related activity are believed to be nearly
ubiquitous among stars of solar and lower  mass, it is less certain
that magnetism is a common characteristic of stars more massive
than the Sun.  To date, definitive detections of fields in stars with
masses $\gta 1.5\ M_\odot$ have, for the most part, been made for objects having
anomalous chemical abundances (e.g., the chemically peculiar A and B stars;
Landstreet 1992; Donati 1998).  Recently, however, observations of cyclic
variability in the properties of winds from luminous OB stars
have been interpreted as evidence for the presence of large-scale
magnetic fields in the surface layers and atmospheres of these objects
(see, e.g., Kaper \& Henrichs 1994; Kaper et al. 1997; Prinja 1998).
These inferences have been bolstered by the unambiguous measurement of 
a weak ($\sim 360$ G) field in the chemically normal B1 IIIe star
$\beta$ Cephei (Donati et al. 2001).  These results suggest that
magnetic fields of moderate strength might be more prevalent among hot
stars than had previously been thought (see Charbonneau \& MacGregor 
2001, hereafter Paper I, for a more complete discussion).

At the present time, the origin of magnetism in massive stars is not well
understood.  It is unclear whether the inferred fields are fossil in nature, 
or are instead the product of dynamo activity in the stellar interior (see, e.g., 
Parker 1979).  The latter explanation, if appropriate, raises additional interesting 
questions concerning the site of magnetic field generation inside hot stars.
For stars having spectral types O and B, convection occurs in the innermost
portion of the core, not in the outer envelope as in low-mass stars.  Because 
convection is thought to be necessary in order that field regeneration via the 
so-called $\alpha$-effect take place, it follows that a hot star dynamo should be 
located deep within the stellar interior.  Within the context of mean-field
electrodynamics, the properties of kinematic core dynamo models for early-type 
stars have been investigated by Levy \& Rose (1974) and Sch\"ussler \& P\"ahler 
(1978), and, more recently, in Paper I.

If the magnetic field of a hot star is produced by dynamo action 
in the convective core, then a mechanism for transporting the field 
to the stellar surface must be identified.  The finite electrical conductivity
of the envelope leads to the outward diffusion of any fields contained therein,
but only over an extended period of time.  Estimates indicate that for stars
more massive than a few solar masses, the resistive diffusion time across the
radiative interior exceeds the main sequence lifetime $t_{MS}$ (Sch\"ussler \& 
P\"ahler 1978).  Another possibility is that dynamo fields are advected from the
core to the surface by rotation-induced meridional circulation (see Paper I).
For a star of mass $M_*$ and equatorial rotation speed $v_{rot}$, a rough
estimate for the circulation time $t_c$ as a fraction of $t_{MS}$ is
$(t_c/t_{MS}) \approx 9 \times 10^2\ (M_*/M_\odot)^{0.8}\ (v_{rot} /
1\ {\rm km\ s}^{-1})^{-2}$ (Kippenhahn \& Weigert 1994).  In the case of
a star with $M_* = 10\ M_\odot$ and $v_{rot} = 150$ km s$^{-1}$,
$(t_c / t_{MS}) \approx 0.25$, indicating that even for relatively rapidly
rotating stars, a time $\sim t_{MS}$ is required to bring the field produced
by the dynamo to the surface by this mechanism.  In addition, the results
of Paper I suggest that the circulatory flow generated by extremely rapid
rotation is capable of interfering with the operation of a core dynamo.

Alternatively, 
in the Sun, magnetic flux emerges at the photosphere in the form of fibrils
or flux tubes.  The field is thought to assume this form in or near
the dynamo domain at the bottom of the convective zone.  The subsequent
rise of a flux tube to the solar surface is driven by buoyancy, a
consequence of the reduced density inside a tube that is in mechanical and
thermal equilibrium with the surrounding, adiabatically stratified,
field-free gas.  If the formation of flux tubes is a process that is not
specific to fields in the Sun, then a dynamo-generated field at the
base of the extended radiative envelope of a hot star might also develop
fibril structure.  In this case, the buoyant force might likewise enable flux
from deep in the interior to reach the surface in a time that is shorter than
evolutionary time scales.  An important distinction, however, is that
unlike the the solar convection zone, the stratification within the
radiative interior of a hot star is sub-adiabatic.  The motion of buoyant
magnetic elements in a thermodynamic environment like that in the envelope of
an upper main sequence star has been studied by Gurm \& Wentzel (1967)
and Moss (1989).

In the present paper, we adopt the premise that hydromagnetic dynamo activity
takes place inside the convective core of a rotating massive star, in a manner
similar to that described in Paper I.  We furthermore presume that the fields
so-produced naturally form into discrete, toroidal flux tubes that remain in 
total pressure equilibrium with the external, unmagnetized stellar interior at
all times.  Using this conceptual framework in concert with the physical model
described in \S2, we determine the time-dependent position, velocity, and 
thermodynamic properties of a buoyant flux ring that begins its outward motion
from a specified location on the core boundary.  Results pertaining to the dynamics
and rise times of rings with a variety of initial field strengths and cross-sectional 
radii are presented in \S3.  Our findings regarding the dynamical behavior of
magnetic rings in a radiative environment and the efficacy of buoyancy as a
flux transport mechanism in hot stars are summarized in \S4.

\section{Model}
To investigate the buoyant transport of magnetic flux through the
radiative interior of a hot star, we consider a spherical star of 
mass $M_*$ and radius $R_*$ that contains a central, convective
core of mass $M_c\ (<M_*)$ and radius $R_c\ (<R_*)$.  The star
is assumed to rotate uniformly with angular velocity $\Omega$
around an axis that coincides with the polar axis of a spherical
coordinate system $(r, \theta, \phi)$.  In the radiative region external
to the core, we study the dynamics of an isolated concentration of azimuthally
directed magnetic field (i.e., a flux tube) that at any given time
takes the form of a circular ring, symmetric about the stellar rotation 
axis.  A similar configuration has been utilized by Choudhuri \& Gilman
(1987) to examine the influence of rotation on the rise of flux tubes
in the solar convection zone.  We assume that the cross-sectional
profile of the tube is a circle whose radius $a$ is much smaller
than either the radius of curvature of the ring or the local scale
height of the surrounding, unmagnetized, radiative envelope. The
assumed geometry for the rising flux tube is illustrated in Figure 1.
These approximations are here used to study the dynamics for the rise
of magnetic flux to the stellar surface, and for comparisons of 
rise times with the main sequence lifetime of the star.

As it is described above, the magnetic ring satisfies the criteria for
applicability of the thin flux tube approximation (e.g., Defouw 1976;
Spruit 1981), and its internal properties can be reasonably taken
to be uniform throughout.  The ring is assumed to contain material
with mass density $\rho$, so that its total mass is $m = \rho V$, where
$V = 2 \pi r\ {\rm sin}\ \theta \cdot \pi a^2$ is the ring volume.
It follows from conservation of mass that if a ring with cross-sectional
radius $a_0$ is released from position $(r_0, \theta_0)$ at time $t=0$,
its density at some $t\ (> 0)$ is
\begin{equation}
\rho = \rho_0 \left( {a_0 \over a} \right)^2
\left( {r_0\ {\rm sin}\ \theta_0 \over r\ {\rm sin}\ \theta}
\right),
\end{equation}
where $a$ and $(r, \theta)$ are the tube radius and position at the
later time.  Likewise, the flux ring is presumed to be untwisted and to
contain a toroidal magnetic field of the form ${\bf B}=B\ {\bf e}_\phi$
with $B$ a constant.  Conservation of magnetic flux then implies that $B$
is related to its initial value $B_0$ through the relation

\begin{equation}
B=B_0 \left( {a_0 \over a} \right)^2.
\end{equation}
A further consequence of the thin flux tube approximation is that
because the time required for a fast magnetosonic wave to transit
the tube cross section is short, the total pressure inside the ring
can be assumed to instantaneously equilibrate with that of the
surrounding stellar material.  The quantitative expression of this lateral
mechanical equilibrium condition is 
\begin{equation}
p_e = p + {B^2 \over 8 \pi},
\end{equation}
where $p = \rho k T / \mu$ is the gas pressure and the subscript $e$
refers to conditions in the external, field-free medium on the periphery
of the tube.  Note that by using the ideal gas law for $p$ and $p_e$,
we are neglecting the radiative component of the total pressure.  This
omission restricts our analysis to stars less massive than about
$10\ M_\odot$, for which radiation pressure is a minor contributor to 
the support of the deep stellar interior.  In addition, we assume that
the mean molecular weight $\mu$ has the same constant value throughout
the star.  Our model thus describes conditions within a young, chemically
homogeneous, main sequence star; it does not apply at later evolutionary
stages when a flux tube that originates in the vicinity of the
core-envelope interface may contain chemically processed material.

In a frame of reference that rotates with the stellar angular velocity,
the uniform flux ring moves coherently in response to the axisymmetric
forces applied to it; that is, each mass element within it has the same 
velocity
\begin{equation}
{\bf u} = u_r\ {\bf e}_r + u_\theta\ {\bf e}_\theta + u_\phi\ {\bf e}_\phi
= {{\rm d} r \over {\rm d} t}\ {\bf e}_r
+ r\ {{\rm d} \theta \over {\rm d} t}\ {\bf e}_\theta
+ r\ {\rm sin}\ \theta\ {{\rm d} \phi \over {\rm d} t}\ {\bf e}_\phi.
\end{equation}
The velocity components $(u_r, u_\theta, u_\phi)$ and ring position $(r, \theta)$ 
can be obtained as functions of time by integration of the ring equation of motion,
\begin{displaymath}
\hspace{-2.0 cm}{{\rm d} {\bf u} \over {\rm d} t} + 2\ {\bf \Omega} \times {\bf u}  =
\left( {\rho_e - \rho \over \rho_e + \rho} \right)
\left[ {G M(r) \over r^2}\ {\bf e}_r - \Omega^2\ r\ ({\rm sin}^2\ \theta\ 
{\bf e}_r
+ {\rm sin}\ \theta\ {\rm cos}\ \theta\ {\bf e}_\theta ) \right] 
\end{displaymath} 
\begin{equation}
\hspace{+4.0 cm} - {B^2 \over 4 \pi r\ (\rho_e + \rho) }\ ({\bf e}_r + 
{\rm cot}~ \theta\    {\bf e}_\theta )
-C_D \left( {\rho_e \over \rho_e + \rho} \right)
{ | {\bf u}_\bot |\ {\bf u}_\bot \over \pi a}, 
\end{equation}
where $M(r)$ is the internal mass distribution of the star, $C_D$ is
the drag coefficient, and ${\bf u}_\bot \equiv u_r\ {\bf e}_r +
u_\theta\ {\bf e}_\theta$ is the transverse velocity.  The individual
components of the inertial and Coriolis terms on the left-hand side
of equation (5) have the explicit forms
\begin{equation}
\hspace{-4cm}\left( {{\rm d} {\bf u} \over {\rm d} t}
+ 2\ {\bf \Omega} \times {\bf u} \right)_r      = 
{ {\rm d} u_r \over {\rm d} t} - { ( u_\theta ^2 + u_\phi ^2 )
\over r} - 2\ \Omega\ u_\phi\ {\rm sin}\ \theta,\\[.1cm]
\eqnum{6a}
\end{equation}
\begin{equation}
\hspace{-2.7cm}\left( {{\rm d} {\bf u} \over {\rm d} t}
+ 2\ {\bf \Omega} \times {\bf u} \right)_\theta =
{ {\rm d} u_\theta \over {\rm d} t} + {u_r u_\theta \over r}
- { u_\phi ^2 \over r}\ {\rm cot}\ \theta
- 2\ \Omega\ u_\phi\ {\rm cos}\ \theta, \\[.1cm]
\eqnum{6b}
\end{equation}
\begin{equation}
\left( {{\rm d} {\bf u} \over {\rm d} t}
+ 2\ {\bf \Omega} \times {\bf u} \right)_\phi  =
{ {\rm d} u_\phi \over {\rm d} t} + {u_r u_\phi \over r}
+ {u_\theta u_\phi \over r}\ {\rm cot}\ \theta
+ 2\ \Omega\ (u_r\ {\rm sin}\ \theta
+ u_\theta\ {\rm cos}\ \theta).
\eqnum{6c}
\end{equation}
\setcounter{equation}{6}

The equation of motion (5) and its many variations have been used to treat
the dynamics of buoyant flux tubes in the Sun's convective envelope by
numerous authors, including Choudhuri \& Gilman (1987), Choudhuri (1989),
Moreno-Insertis, Sch\"ussler, \& Ferriz-Mas (1992), Cheng (1992),
Fan, Fisher, \& DeLuca (1993), Ferriz-Mas \& Sch\"ussler (1993, 1994),
Fan, Fisher, \& McClymont (1994), Caligari, Moreno-Insertis,
\& Sch\"ussler (1995), and Fan \& Fisher (1996), among others.
The three terms on the right-hand side of equation (5) describe,
respectively, the accelerations produced by: ($i$) the buoyant force,
including the effect of the centrifugal reduction of the local gravitational
acceleration; ($ii$) the magnetic tension force, arising in response to
outward, stretching motion of the ring in the plane perpendicular to the
rotation axis of the star; and ($iii$), the aerodynamic drag force, derived
from that exerted on a straight, circular cylinder immersed in a steady flow
with upstream velocity directed perpendicular to the axis of symmetry
(Goldstein 1938).  The appearance of the factor $(\rho_e + \rho)^{-1}$
rather than $\rho^{-1}$ in the expressions for each of the accelerations
reflects the fact that we have made provision for the so-called virtual (or
enhanced) inertia of the ring, the hydrodynamical resistance to acceleration
experienced by a rigid body that moves through a fluid (see, e.g., Batchelor
1967).  Considerable discussion has been devoted to the reasons for
including (or not including) this effect in the equation describing the
balance of forces acting on a flux tube (see, e.g., Moreno-Insertis,
Sch\"ussler, \& Ferriz-Mas, 1996; Fan \& Fisher 1996, and references
therein).  As will become clear in subsequent sections, the overall picture
of flux ring dynamics that emerges from solutions of equation (5) depends
little on the presence or absence of virtual inertia in any of the terms.

The difference between the nature of the energy balance that prevails within 
the envelope of a massive star and that within the convection zone of the Sun 
leads to significant differences between the dynamics of flux tubes located 
in the two regions.  The outer portion of the solar interior is adiabatically
stratified, a consequence of efficient convective energy transport therein.
Because of this, an adiabatic flux tube that is initially buoyant will remain
so during the course of its rise toward the surface.  Alternatively, the envelope
of a hot star is sub-adiabatically stratified, a consequence of the radiative
equilibrium conditions that exist throughout it.  As a result, the density deficit
of an initially buoyant, adiabatic tube will diminish as it rises, causing
the upward buoyant acceleration to decrease as well.  When adiabatic cooling
causes the temperature contrast of the tube relative to its surroundings to
attain the value $(T_e -T)/T_e = (1 + \beta )^{-1}$ where $\beta = (8 \pi p
/ B^2)$, $\rho = \rho_e$, and the buoyancy of the tube is reduced to zero.

The discussion of the preceding paragraph indicates that within the envelope
of a massive star, the non-adiabatic, radiative interaction between a flux
ring and its environment will play an important role in determining the
magnitude of the buoyant acceleration the ring experiences, and thus, the
time required for it to emerge at the surface of the star (see, e.g., Parker
1975; Moreno-Insertis 1983; Fan \& Fisher 1996).  In the present model, the
heating associated with the diffusion of radiation from the external medium
into the tube is described by the thermal energy equation (see Appendix A),
\begin{equation}
{ {\rm d} S \over {\rm d} t} =
{32 \over 3}\ { (\gamma -1)\ \sigma\ T_e^4 \over
\kappa_e\ \rho_e\ p\ a^2}\ {\Delta T \over T_e} \equiv q, 
\end{equation}
where $\Delta T = (T_e - T)$, and $S$ is the entropy-like quantity
\begin{equation}
S = {\rm ln} \left[ \left( {p \over p_0} \right)
\left( {\rho_0 \over \rho} \right) ^\gamma \right]. 
\end{equation}
In equations (7) and (8), $\gamma$ is the ratio of specific heats, $\sigma$
is the Stefan-Boltzmann constant, and $\kappa_e$ is the Rosseland mean opacity
in the material surrounding the tube.  In the absence of radiative heating
(i.e., when $q=0$), equation (7) indicates that $S$ is constant, and the
tube behaves adiabatically.

A general question concerning the flux tube approach is how far we can
consider the flux tube to be thin in the thermal sense. We assume the
heating that arises from the inward diffusion of radiation affects the tube
interior uniformly. In reality, the fact that the heating time scale and the
tube radius are both finite implies a differential response by the tube to
the heat input. In particular, if the time scale for the dynamical
adjustment of the tube ($\sim$ the sound crossing time) is much shorter than
the diffusion heating time scale, then the temperature in the outermost
portion of the tube can become elevated relative to the center, and the
resultant enhancement of the buoyant acceleration can cause a deformation of
the tube. Unless checked by some additional confining influence (e.g.,
twisted flux tube fields), this tendency might ultimately lead to the
destruction of the flux concentration.  While quantitative treatment of this
effect is beyond the scope of the present exploratory model calculations,
its potential occurence underscores the desirability of performing 2D MHD
simulations of flux tubes in a radiative enviroment.

We use the model consisting of equations (1)-(8) to determine
the motion of an initially buoyant flux ring in the radiative envelope of
a hot star.  The distributions of pressure ($p_e$), density ($\rho_e$), and
temperature ($T_e$) are taken from a spherical, non-rotating stellar interior
model for $M_* = 9\ M_\odot$, kindly calculated for us by S. Jackson. 
As described in Appendix B,
in order to facilitate numerical computations, we have derived an approximate,
analytic representation of this model by assuming that $p_e$, $\rho_e$,
and $T_e$ are related polytropically.  For the purpose of these simulations, 
the star is assumed to rotate rigidly with a prescribed angular velocity
$\Omega$ that is sufficiently slow that departures from sphericity can be 
neglected. 

At time $t=0$, a toroidal flux ring with specified values of the cross-sectional
radius $a_0$ and plasma beta $\beta_0\ (=8 \pi p_0 /B_0^2)$ is released 
from the outer boundary of the convective core ($r_0 = R_c$) at co-latitude 
$\theta_0$ (latitude $\lambda_0$).  We suppose that the ring is initially
in thermal equilibrium with the ambient stellar material, $T_0 = T_{e0}$.
Application of the mechanical equilibrium condition (3) then yields
$p_0 = p_{e0}\ \beta_0 /(1 + \beta_0)$ and $\rho_0 = \rho_{e0}\ 
\beta_0/(1 + \beta_0)$, so that $(\rho_{e0} - \rho_0) >0$ and the ring
is buoyant.  The initial field strength $B_0$ and Alfv\'en
speed $u_{A0}$ in the ring are given in terms of $\beta_0$ and the
external properties at the starting position according to $B_0 = [8 \pi p_{e0} 
/(1 + \beta_0)]^{1/2}$ and $u_{A0} = [2 p_{e0} / (\beta_0 \rho_{e0})]^{1/2}$.

The velocity components ($u_r, u_\theta, u_\phi$) are obtained as
functions of $t$ by numerical integration of the three components of
equation (5); the position ($r, \theta$) of the axisymmetric ring 
follows by integrating the relations ${\rm d} r /{\rm d} t = u_r$,
${\rm d} \theta / {\rm d} t = u_\theta /r$ given in equation (4).
In addition, we simultaneously integrate the thermal energy equation (7)
to obtain $S$, which enables us to update the tube properties at 
each time step.  In particular, from the definition given in equation (8),
the pressure $p$ within the tube can be expressed as

\begin{equation}
p = p_0\ e^S\ \left( {\rho \over \rho_0} \right)^\gamma.
\end{equation}
Similarly, if equation (1) is rewritten in the form
\begin{equation}
\left( {a \over a_0} \right)^2 = \left( {\rho_0 \over \rho} \right)
\left( {r_0\ {\rm sin}\ \theta_0 \over r\ {\rm sin}\ \theta} \right),
\end{equation}
then substitution in equation (2) yields
\begin{equation}
B = B_0 \left( {\rho \over \rho_0} \right)
\left( {r\ {\rm sin}\ \theta \over r_0\ {\rm sin}\ \theta_0} \right).
\end{equation}
This result, together with equation (9), permits the pressure equilibrium
condition (3) to be recast as
\begin{equation}
e^S \left( {\rho \over \rho_0} \right)^\gamma +
{1 \over \beta_0}
\left( {r\ {\rm sin}\ \theta \over r_0\ {\rm sin}\ \theta_0} \right)^2
\left( {\rho \over \rho_0} \right)^2
-\left( 1 + {1 \over \beta_0} \right)
\left( {p_e \over p_{e0}} \right) = 0, 
\end{equation}
which is solved to obtain $\rho / \rho_0$, given the values of
$S$ and $p_e$ that prevail at a particular time $t$ and location
$(r, \theta)$.  Equations (9), (10), and (11) are then used to derive
$p$, $a$, and $B$, and $T$ is determined via the relation $T /T_0
= (p / p_0) (\rho_0 / \rho)$.

\section{Results}

We have used the thin flux tube physical model and solution method
outlined in the preceding section to investigate the dynamical behavior of
magnetic flux rings with $10^4 \leq \beta_0 \leq 10^7$ and $10^{-4} \leq
(a_0 / h_{e0}) \leq 10^{-2}$, where $h_{e0}$ ($= 2.65 \times 10^{10}$ cm;
see Appendix B) is the pressure scale height in the stellar interior at the
starting radius $r_0 = R_c$.  The maximum and minimum values adopted for the
quantity $\beta_0$ correspond to toroidal fields with strengths $B_0 = 1.82
\times 10^5$ G and $5.77 \times 10^6$ G, respectively.  In the computational
results described below, the parameters $\gamma$, $C_D$, and $\Omega$ were
assigned the values, $\gamma = 5/3$, $C_D =1$, and $\Omega = 5.88
\times 10^{-5}$ s$^{-1}$, the latter value representing a surface
equatorial rotation speed of about 150 km s$^{-1}$.  We have
examined the motion of buoyant flux rings following release from
the core boundary at initial latitudes in the range $0^\circ \leq
\lambda_0 \leq 45^\circ$ (i.e., $\pi /4 \leq \theta_0 \leq  \pi /2$).
Because the dynamics of rings in the equatorial plane ($\lambda_0
= 0^\circ$) differ from those of rings with $\lambda_0 > 0^\circ$,
we treat the two cases separately in the ensuing discussion.

\subsection{Flux Rings in the Equatorial Plane}

Inspection of the equation of motion (5) reveals that the meridional
component of each force acting on a flux ring that is initially at
rest (${\bf u} = 0$) in the equatorial plane ($\theta = \pi / 2$)
vanishes identically.  The subsequent motion of such a ring therefore
remains in this plane, and is governed by just the $r$- and $\phi$-components
of equation (5).  Some of the properties of configurations of this kind
are illustrated in Figures 2 and 3, which contain results pertaining to
rings with $\beta_0 = 10^4$.  The pairs of panels that make up the three
rows of Figure 2 show, respectively, the time evolution of: ($i$) the radial
position of the ring, $(\Delta r / r_0) = (r - r_0)/r_0$; ($ii$) the velocity 
components $(u_r, u_\theta, u_\phi)$ in units of the Alfv\'en speed $u_{A0}\ 
(= 7.49 \times 10^5$ cm s$^{-1}$) in the ring at $t = 0$; and ($iii$),
the accelerations produced by each of the forces acting on the ring,
in units of $u_{A0}^2 / R_*$.  In each panel, time is measured in units
of the Alfv\'en time $t_A$, defined as $t_A \equiv R_*/u_{A0}\ (=3.42 \times
10^5$ s $\approx 4$ days).

The dynamical evolution of a flux ring with initial cross-sectional
radius $a_0 = 10^{-4}\ h_{e0}$ is shown in panels (A), (C), and (E) of
Figure 2.  Note that in this case, the outward expansion of the ring 
ceases after $t \approx t_A$, its radial position
remaining fixed at later times.  In the course of evolving toward this
apparent equilibrium configuration, the ring experiences a brief, initial
period of accelerated motion in the $+{\bf e}_r$-direction, followed by radial
deceleration to a state of rest in which $u_r = 0$ but $u_\phi \neq 0$.
Examination of the ring force balance indicates that the initial expansion
is driven (as expected) by buoyancy, while being opposed by the slightly
smaller, inward-directed, magnetic tension force.  The aerodynamic drag
force, although small in magnitude, contributes to the damping of the
oscillatory motions that are induced by the interplay between the larger
buoyant and magnetic forces.

The flux ring is assumed to corotate with the stellar interior at $t = 0$, 
so that its initial azimuthal velocity is $u_{\phi 0}= 0$ in the reference
frame that rotates with angular velocity $\Omega$.  For $t > 0$, the
azimuthal motion of the ring is such that the angular momentum of the
material contained within it is conserved.  The buoyancy-driven increase
in the radial position of the ring is therefore accompanied by a
decrease in the rotation rate of the ring material.  Hence, in the
frame of reference that corotates with the star, the azimuthal velocity
of the ring is in the $-{\bf e}_\phi$-direction, as can be seen in
panel (C).  According to equation (6$a$), this azimuthal motion
produces an inward-directed Coriolis force that grows with increasing
$\Delta r$ until its magnitude, combined with that of the magnetic
tension force, is sufficient to balance the buoyant force and prevent
further radial expansion of the ring.  A similar behavior has been
observed in connection with models of equatorial magnetic flux rings
located in the radiative layers just below the base of the solar convection
zone (Moreno-Insertis, Sch\"ussler, and Ferriz-Mas 1992; Ferriz-Mas 1996).

In panels (A) and (C) of Figure 3, we show how the thermodynamic state
of a ring with $\beta_0 = 10^4$ and $a_0 = 10^{-4}\ h_{e0}$ evolves
over time.  Panel (A) shows the normalized differences between the pressure, 
density, and temperature in the tube and in the material surrounding it
(i.e., $\Delta p/p_e$, $\Delta \rho/\rho_e$, and $\Delta T/T_e$,
respectively, where for any quantity $Q$, $\Delta Q \equiv Q_e -Q$) as
functions of time.  Beginning from a state in which $T_0 = T_{e0}$, 
the ring cools slightly relative to its surroundings during the earliest 
and most rapid portion of its limited buoyant ascent.  As its outward
progress stalls, radiative heating acts to restore the condition of
thermal equilibrium between the ring and its external environment.
In panel (C), we display the time evolution of the quantities $S$ and $q$
(see equations [7] and [8]), along with the corresponding histories of the 
instantaneous heating and rise times, defined as $t_H \equiv q^{-1} $
(see Appendix A) and $t_R \equiv |h_e / u_r|$, respectively.  Note that $S$ 
increases as the tube is radiatively heated, and approaches a constant value 
as $u_r \rightarrow 0$, $T \rightarrow T_e$, and both $t_H$ and $t_R$ become 
large.

To further explore the nature of the equilibrium seen in Figures 2 and 3
for flux rings in the equatorial plane, we write the $\phi$-component of 
the equation of motion (5) in the form
\begin{equation}
r {{\rm d} u_\phi \over {\rm d} t} + u_r u_\phi
+ 2 u_r \Omega r 
= {{\rm d} \over {\rm d} t} \left( r u_\phi + \Omega r^2 \right)
=0, 
\end{equation}
where we have assumed $u_\theta = 0$ for $\theta = \pi / 2$, and have used
the fact that $u_r = {\rm d} r / {\rm d} t$.  Equation (13) implies that
the specific angular momentum $(r u_\phi + \Omega r^2)$ has a constant value
which, after recalling the initial condition $u_{\phi 0} = 0$, is readily
established to be $\Omega r_0^2$.  This quantity can then be used to
evaluate the azimuthal velocity component of the ring, yielding the result
\begin{equation}
u_\phi = \left( {\Omega r_0^2 \over r} \right) - \Omega r. 
\end{equation}

Similar considerations applied to the $r$-component of equation (5)
lead to the equilibrium condition
\begin{equation}
\left( {\rho_e - \rho \over \rho_e + \rho} \right)
\left[ {G M(r) \over r^2} - \Omega^2\ r \right]
+ {u_\phi^2 \over r} + 2\ \Omega\ u_\phi
- {B^2 \over 4 \pi r\ (\rho_e + \rho)} = 0, 
\end{equation}
where we have set $u_r = ({\rm d} u_r / {\rm d} t) = 0$.  Equation (15)
expresses the fact that the radial equilibrium is characterized by a
balance between the outward-directed buoyancy and centrifugal forces
and the inward-directed Coriolis and magnetic tension forces.  Making
the substitution $(r/r_0) = 1 + (\Delta r/r_0)$ and retaining only the
lowest order terms in $(\Delta r/r_0)$ ($\ll 1$), we obtain
\begin{equation}
{\Delta r \over r_0} \approx {1 \over 8 (\Omega r_0)^2}
\left( {G M_c \over \beta_0 r_0} - u_{A0}^2 \right)
= {1 \over 8 \beta_0 (\Omega r_0)^2}
\left( {G M_c \over r_0} - {2 p_{e0} \over \rho_{e0}} \right),
\end{equation}
where equation (14) and the approximation $\rho_e \approx \rho\
[ 1 + (1/\beta_0)]$ have been used in simplifying equation (15).

As is evident from equation (16), for a given stellar model, the
equilibrium position of an equatorial flux ring depends only on the
value of the parameter $\beta_0$; in the case of the present model, 
$(\Delta r / r_0) \approx 6 \times 10^{-4}\ (\beta_0 /10^4)^{-1}$.
This dependence has been confirmed by examination of solutions
corresponding to a variety of input parameter values, including
the particular example shown in panels (B), (D), and (F) of Figure 2,
and in panels (B) and (D) of Figure 3.  For this solution, $\beta_0
= 10^4$ as in the case of the ring considered above, but $a_0 =
10^{-3}\ h_{e0}$.  Because this initial cross-sectional radius is
a factor of 10 larger than the previous value, the radiative heating
rate $q$ and the acceleration produced by the aerodynamic drag
force are smaller by factors of 100 and 10, respectively (see
equations [5] and [7]).  As can be seen in the relevant portions of
Figures 2 and 3, a consequence of these reductions is that the ring
is subject to vigorous buoyancy oscillations in the radial direction.
Oscillations of a similar kind in toroidal flux tubes inside non-rotating
stars have been studied by Spruit \& van Ballegooijen (1982), and
inside rotating stars by van Ballegooijen (1983), Moreno-Insertis,
Sch\"ussler, \& Ferriz-Mas (1992), and Ferriz-Mas \& Sch\"ussler
(1993).  In the present example, the enhanced tendency toward
oscillatory motion is directly attributable to modifications in the
buoyant acceleration of the ring; these changes are themselves a
result of the altered thermodynamic state of the ring material,
produced by the diminished rate of radiative heating.  The build-up
in the magnitude of the Coriolis force, together with the action of
the drag force, cause the amplitude of these oscillations to decrease
over time, leaving the ring in the same equilibrium  configuration
that obtained in the case of the $a_0 = 10^{-4}\ h_{e0}$ ring studied
above.

These results suggest that the equilibrium of a flux ring located in
the equatorial plane is stable with respect to small amplitude, radial
displacements from the position given by equation (16).  This conjecture
has been verified by a series of numerical experiments in which an
additional, radially directed force of specified amplitude is applied 
to the ring for a brief time interval after it has assumed its equilibrium 
position.  In all cases, the ring is initially displaced in radius, but
is quickly restored to its original location following cessation of the
imposed forcing.  Such behavior is not observed in response to the
application of latitudinal forcing of the same type.  In all of these
cases, a small displacement out of the equatorial plane is accompanied 
by the development of a $\theta$-component of the magnetic tension force
that accelerates the ring toward the pole.  This is similar to the
dynamical evolution of the so-called poleward-slip instability of equatorial
flux rings in the radiative layers beneath the solar convection zone (see, e.g., 
Moreno-Insertis, Sch\"ussler, \& Ferriz-Mas 1992, and references therein).
For the conditions considered in the present paper, the perturbed force balance 
is such that a ring initially contained within the equatorial plane moves over 
time to higher latitudes and larger radii.  Once out of the equatorial plane, 
the dynamics of such a ring is found to be identical to that of one with
$\theta_0 < \pi /2$, and is discussed in \S3.2 below.

\subsection{Flux Rings Outside of the Equatorial Plane}

We now consider the motion of toroidal magnetic flux tubes having initial
positions $r_0 = R_c$ with $0 < \theta_0 < \pi/2$.  We focus on the properties
of rings with $\theta_0 = \pi/3$ (i.e., initial latitude $\lambda_0 = 30^\circ$)
since the behavior in this case is representative of that displayed by all rings
having starting locations along the core-envelope interface, out of the equatorial
plane.  In Figure 4, we show the trajectory that a ring with $\beta_0 = 10^4$
and $a_0 = 10^{-4}\ h_{e0}$ follows during the time interval between its release 
and $t = 10^3\ t_A \approx 10.85$ years.  As can be seen in panel (A),
the edge of the ring traces a path that is nearly parallel to the stellar
rotation axis over this period, extending from $r_0 = 0.232\ R_*$ and $\lambda_0
= 30^\circ$ to $r = 0.64\ R_*$ and $\lambda = 72^\circ$.  In panel (B), we have
magnified the horizontal scale in order to make the meridional motion of the
ring more apparent.  There it can be seen that for a time $t \approx
5\ t_A$ after the start of the calculation, the distance $r\ {\rm sin}\  
\theta$ of the ring from the stellar rotation axis decreases as the ring
moves toward higher latitudes along the periphery of the convective core.  At 
later times, this moment arm length approaches its initial size, while the 
height $r\ {\rm cos}\ \theta$ of the ring above the equatorial plane steadily 
increases.

Additional information pertaining to the evolving dynamical and thermodynamical
properties of the flux ring under consideration is presented in Figure 5.
Inspection of panels (A) and (B) reveals that after a brief, initial period
of dynamical readjustment, the ring moves continuously in the direction of higher 
latitudes (i.e., the $-{\bf e}_\theta$-direction) and larger radii.  This is 
unlike the behavior of rings with $\lambda_0 = 0^\circ$ which, in the absence
of suitable perturbations, do not depart from the equatorial plane as they
evolve toward a final equilibrium state with $u_r = u_\theta = 0$ (see \S3.1).
In the present case, although its upward progress slows considerably at later
times, the ring never attains an equilibrium in which the forces acting in
the ${\bf e}_r$- and ${\bf e}_\theta$-directions balance to produce a state 
of rest.

Details concerning the radial and meridional dynamics of the flux ring
are provided in panels (C) and (D) of Figure 5.  When the ring is
released at $t = 0$, it experiences a net radial acceleration that is $>0$,
resulting from the fact that the outward, buoyant force is larger than the
inward, radial component of the tension force.  In the ${\bf e}_\theta$-direction,
the tension force is initially unopposed by any other force components, leading
to a net meridional acceleration that is $<0$.  The ring is therefore pulled
toward the pole, and the horizontal distance $r\ {\rm sin}\ 
\theta$ between it and the rotation axis decreases (see Figure 4).  This change
in position has an important consequence for the subsequent motion of the ring. 
In order to conserve angular momentum, the azimuthal velocity $u_\phi$ increases
as the moment arm of the ring material becomes smaller (see panel [B]).
Associated with this spin-up are substantial increases in the magnitudes of the 
$r$- and $\theta$-components of the Coriolis force, both of which exert
considerable influence over the dynamical evolution of the ring.

Using a procedure similar to that adopted in the case of an equatorial flux ring, 
the azimuthal component of the equation of motion (5) for a ring at higher
latitudes can, after some manipulation, be written as
\begin{equation}
{{\rm d} \over {\rm d} t} \left( r\ {\rm sin}\ \theta\ u_\phi
+ \Omega r^2\ {\rm sin}^2\ \theta \right) =0, 
\end{equation}
from which it immediately follows that
\begin{equation}
u_\phi = \left( {\Omega r_0^2\ {\rm sin}^2\ \theta_0 \over
r\ {\rm sin}\ \theta} \right) - \Omega r\ {\rm sin}\ \theta. 
\end{equation}
Note that in this case, because the initial radial and meridional motion of 
the ring corresponds to decreasing $r\ {\rm sin}\ \theta$, the sense of the
azimuthal motion is such that $u_\phi > 0$.  Hence, the accelerations
produced by the Coriolis force in the ${\bf e}_r$- and ${\bf e}_\theta$-directions,
$2\ \Omega\ u_\phi\ {\rm sin}\ \theta$ and $2\ \Omega\ u_\phi\ {\rm cos}\ 
\theta$, respectively, are both $>0$.  As a result, the tension-induced,
poleward movement of the ring at early times begins to slow when the sum of 
the $\theta$-components of the Coriolis and drag forces becomes large enough
to change the sign of the net meridional acceleration, making it $>0$.
Likewise, the acceleration supplied by buoyancy and the radial Coriolis force component
causes $u_r$ to grow until $u_r = -u_\theta\ {\rm cot}\ \theta$, at which point contraction
of the ring toward the rotation axis ceases; at later times, $u_r > -u_\theta\
{\rm cot}\ \theta$, and the ring expands.  The resulting increase in
$r\ {\rm sin}\ \theta$ leads to a gradual reduction in $u_\phi$, and leaves
the ring in a state of near balance between the Coriolis and tension forces
in both the ${\bf e}_r$- and ${\bf e}_\theta$-directions.

The lack of equilibria for flux rings that originate at latitudes
$\lambda_0 > 0^\circ$ can be understood by examining the $r$- and 
$\theta$-components of the equation of motion (5).  Assuming that a
stationary state with $u_r = u_\theta = 0$ exists, the $r$-component can
be written in a form reminiscent of the equilibrium condition (15) for
flux rings in the equatorial plane,
\begin{equation}
\left( {\rho_e - \rho \over \rho_e + \rho} \right)
\left[ {G M(r) \over r^2} - \Omega^2\ r\ {\rm sin}^2\ \theta \right]
+{u_\phi^2 \over r} + 2\ \Omega\ u_\phi\ {\rm sin}\ \theta
-{B^2 \over 4 \pi r\ (\rho_e + \rho)} = 0, 
\end{equation}
while the $\theta$-component becomes
\begin{equation}
- \left( {\rho_e - \rho \over \rho_e + \rho} \right)\ 
\Omega^2\ r\ {\rm sin}^2\ \theta +{u_\phi^2 \over r}
+ 2\ \Omega\ u_\phi\ {\rm sin}\ \theta
+{B^2 \over 4 \pi r\ (\rho_e + \rho)} = 0. 
\end{equation}
Note that were it not for the appearance of the gravitational acceleration
in the radial component, equations (19) and (20) would be identical.  This
correspondence implies that the conditions for force balance in 
the ${\bf e}_r$- and ${\bf e}_\theta$-directions can be simultaneously 
satisfied only if $\rho =\rho_e$. However, as can be seen in panel (E) of 
Figure 5, $\rho < \rho_e$ throughout
the time interval covered by the calculation, indicating that a stationary
equilibrium is not possible in this particular case.

In general, all magnetic flux rings with $\lambda_0 > 0^\circ$ remain slightly 
cooler and less dense than the surrounding material during most of their
ascent, a consequence (in part) of the total pressure balance condition
given by equation (3).  On the basis of equations (19) and (20) then,
stationary equilibria are ruled out for toroidal flux tubes of this kind.
Because $T < T_e$ (see panel [E] of Figure 5), a given
tube is heated by the radiation that diffuses into it from the external medium.
In response to this heating and in order to comply with the pressure equilibrium
condition, the tube expands, becoming less dense as it does so.  Buoyancy
then carries the tube outward to a somewhat larger radius.  Hence, apart
from a brief period at the start of the motion when the buoyant acceleration
is established by the initial conditions, the rate of rise of the tube is
controlled by the rate at which it is radiatively heated.  This behavior is
evident in panel (F) of Figure 5, wherein it can be seen that for much of the
time interval depicted, the heating and rise times vary in concert.  Such
a situation is in contrast to the dynamical and thermodynamical evolution
observed previously for flux rings in the equatorial plane.  The stationary
equilibrium of an equatorial flux ring is characterized by $T = T_e$,
a vanishing radiative heating rate, and a balance between buoyancy and
the sum of the Coriolis and tension forces (see Figures [2] and [3]).

The considerations of the preceding paragraph can be used to derive an
approximate expression for the radial rise speed of a buoyant flux ring.
Interpreting the derivative on the left-hand side of equation (7) as
$({\rm d} S /{\rm d} t) = u_r ({\rm d} S / {\rm d} r)$, the sought-after
component of the tube velocity is
\begin{equation}
u_r = { q \over ({\rm d} S /{\rm d} r)}, 
\end{equation}
where $q$ and $S$ are defined in equations (7) and (8).  To evaluate
$({\rm d} S /{\rm d} r)$, we assume that since $\beta \gg 1$, $p \approx p_e$
and $\rho \approx \rho_e$; then,
\begin{equation}
S \approx {\rm ln} \left[ {p_{e0} \over p_0} 
\left( {\rho_0 \over \rho_{e0}} \right)^\gamma \right]
+ \left( {\alpha - \gamma \over \alpha} \right)
{\rm ln} \left( {p_e \over p_{e0}}\right), 
\end{equation}
where we have made use of the polytrope relation in the form $(\rho_e / \rho_{e0})
= (p_e / p_{e0})^{1/\alpha}$.  Thus,
\begin{equation}
{{\rm d} S \over {\rm d} r} \approx
\left( {\alpha - \gamma \over \alpha} \right) { {\rm d\ ln}\ p_e \over
{\rm d} r} = \left( {\gamma - \alpha \over \alpha} \right)
{ 1 \over h_e}, 
\end{equation}
and
\begin{equation}
u_r \approx \left( {\alpha \over \gamma - \alpha} \right)
q\ h_e, 
\end{equation}
where $h_e = p_e /(\rho_e g)$ is the pressure scale height.  In Figure 6,
we compare the approximation given in equation (24) with the radial velocity
component obtained by numerical solution of the full equation of motion for
$\beta_0 = 10^4$, $a_0 = 10^{-4}\ h_{e0}$, and $\lambda_0 =30^\circ$.  At
times later than $t \approx t_A$, the two results are indistinguishable,
indicating that the rate of ascent is indeed regulated by the radiative
heating of the ring.  In fact, using the definitions of the heating and
rise time scales, equation (24) implies that $t_R \propto t_H$, in accordance
with the behavior seen in panel (F) of Figure 5.

Figures 7 and 8 contain a summary of results from the simulated rise of
a toroidal magnetic flux tube released at latitude $\lambda_0 = 30^\circ$
with $\beta_0 = 10^4$ and $a_0 = 10^{-3}\ h_{e0}$.  As was discussed in \S3.1,
the larger cross-sectional radius of this tube implies that the deceleration
due to aerodynamic drag ($\propto a^{-1} $) and the radiative heating rate 
($\propto a^{-2}$) are both reduced from the magnitudes these quantities
have when $a_0 = 10^{-4}\ h_{e0}$.  Similar to the equatorial flux rings that 
were the subject of that section, in the present case, these reductions are
responsible for engendering oscillatory motion of the tube during the early
stages of its ascent toward the surface.  Such behavior can be seen in
Figure 7, in which (as in Figure 4) two views of the path taken by the
rising tube are given.  In panel (A), it is apparent that the larger tube
is less buoyant; during the time interval $\Delta t = 10^3\ t_A$ between the
first and last points on the trajectory, the radial position of the ring only
increases from $r_0 = 0.232\ R_*$ to $r = 0.275\ R_*$.  On the expanded scale
of panel (B), it is seen that for a time $\approx 0.25\ t_A$ following the
start of the calculation, the ring moves steadily toward the pole, and
thereafter executes latitudinal oscillations that decrease in amplitude as
the distance of the ring from the equatorial plane increases.

Inspection of the relevant panels of Figure 8 indicates that as in the case
of the $a_0 = 10^{-4}\ h_{e0}$ tube considered earlier in this section, the
initial poleward motion is driven by the magnetic tension force.  However, in 
the present case, the tube acquires a higher meridional velocity and approaches
closer to the rotation axis during this movement, a consequence of the fact
that the deceleration due to drag is smaller.  Also, the change in volume
arising from the decrease in $r\ {\rm sin}\ \theta$, together with the
diminshed efficiency of radiative heating, are sufficient to ultimately
make $\Delta \rho < 0$, and to thereby change the sign of the buoyant acceleration
(see panels [C] and [E]).  Note that the the spin-up associated with the
overall contraction of the ring causes both components of the Coriolis
acceleration to increase substantially; the oscillatory behavior seen in
Figure 7 results from the interplay between this acceleration and that
produced by the magnetic tension force.

\subsection{Rise Times}

A primary focus of the present investigation is the buoyant transport of
magnetic flux from the assumed site of its generation in the convective core
to the stellar surface.  Of particular interest is an estimate of the time
required for this transport to take place.  From the definition of $u_r$
given in equation (4), it follows that
\begin{equation}
t_r(r)= \int \limits^r _{r_0} {{\rm d}r \over u_r}, 
\end{equation}
is the time it takes a flux ring to travel outward from $r_0$ to any radius
$r > r_0$. 

We have computed $t_r(r)$ for toroidal flux tubes having
the properties listed in Table 1.  For each case, the radial component of
the tube velocity was determined by numerical integration of the full
equation of motion (5) for a time interval $0 \leq t \leq t_1$, where the
value of $t_1$ for a given solution was chosen to be between $10^3\ t_A$ and
$3 \times 10^4\ t_A$, depending upon the magnitude of $u_r$ for that solution.
For $t > t_1$, $u_r$ was evaluated using an analytic approximation to equation
(24), derived in the following way.
\begin{deluxetable}{cccccc}
\tablecaption{Properties of Magnetic Flux Ring Solutions}
\label{Tab1}
\tablewidth{0pt}
\tablehead{
\colhead{Number} &
\colhead{$\beta_0$} &
\colhead{$a_0/h_{e0}$} &
\colhead{$C_D$} &
\colhead{$B_0$} &
\colhead{$t_r(0.96\ R_*)$} \nl
\colhead{} &
\colhead{} &
\colhead{} &
\colhead{} &
\colhead{(G)} &
\colhead{(years)} }
\startdata
 1 \tablenotemark{ } & $10^4$ & $10^{-4.0}$ & 1 & $5.77 \times 10^6$ & 
      $1.3010 \times 10^5$ \nl
 2 & $10^4$ & $10^{-3.0}$ & 1 & $5.77 \times 10^6$ & $1.3225 \times 10^7$ \nl
 3 & $10^4$ & $10^{-4.0}$ & 0 & $5.77 \times 10^6$ & $1.3009 \times 10^5$ \nl
 4\tablenotemark{a} & $10^4$ & $10^{-4.0}$ & 1 & $5.77 \times 10^6$ & 
      $1.3020 \times 10^5$ \nl
 5 & $10^5$ & $10^{-4.0}$ & 1 & $1.82 \times 10^6$ & $1.3011 \times 10^6$ \nl
 6 & $10^5$ & $10^{-3.5}$ & 1 & $1.82 \times 10^6$ & $1.3024 \times 10^7$ \nl
 7 & $10^5$ & $10^{-3.0}$ & 1 & $1.82 \times 10^6$ & $1.3037 \times 10^8$ \nl
 8 & $10^6$ & $10^{-4.0}$ & 1 & $5.77 \times 10^5$ & $1.3011 \times 10^7$ \nl
 9 & $10^6$ & $10^{-3.0}$ & 0 & $5.77 \times 10^5$ & $1.3013 \times 10^9$ \nl
10 & $10^7$ & $10^{-4.0}$ & 1 & $1.82 \times 10^5$ & $1.3011 \times 10^8$ \nl
\enddata
\tablenotetext{ }{Note$.-$All solutions have $\lambda_0 = 30^\circ$.}
\tablenotetext{a}{Virtual inertia is omitted from solution 4.}
\end{deluxetable}
According to equation (24), $u_r \propto q\ h_e$ at times late enough that any
transient behavior stemming from the adjustment of the tube to the imposed initial 
conditions has disappeared.  The heating rate $q$ depends directly on the
temperature difference $\Delta T$ between the tube and its surroundings; using
the pressure equilibrium condition (3), it can be shown that $(\Delta T / T_e) \approx
B^2/(8 \pi p_e)$, assuming $\rho \approx \rho_e$.  Inserting this result in
the definition of $q$ (see Appendix A), we find that the dependence of $u_r$ on
the physical properties of the tube and the external medium is given by
(see also Gurm \& Wentzel 1967; Parker 1975)
\begin{equation}
u_r \approx {32 \over 3} \left[ {\alpha\ (\gamma -1) \over \gamma - \alpha}
\right] \left( {\sigma\ T_e^4 \over \kappa_e\ \rho_e^2\ g} \right)
\left( {B^2 \over 8 \pi\ p\ a^2} \right). 
\end{equation}
To further simplify this expression, note that as evidenced by Figures 4 and 7,
much of the ascent toward the surface takes place with $r\ {\rm sin}\ \theta
\approx$ constant (i.e., $u_r \approx -u_\theta\ {\rm cot}\ \theta$).  Therefore,
from equations (1) and (2), $\rho\ (\approx \rho_e)\ \propto a^{-2}$ and $B 
\propto a^{-2} \propto \rho_e$, so that $u_r$ varies as
\begin{equation}
u_r \propto {T_e^3 \over \kappa_e\ g}, 
\end{equation}
where we have approximated $p\ a^2 \propto T \approx T_e$.  To illustrate
the validity of equation (27), we have evaluated the constant of proportionality
using the properties of solution 1 (see Table 1) at the time $t = 50\ t_A$.  The
rise speed so-derived is shown as a function of time for $t \geq 50\ t_A$ by
the dotted line in Figure 6.

The results of our rise time computations are depicted in graphical form in
Figure 9.  There we show the time $t_r(r)$ required to reach a given radius 
$r$ within the interval $0.232\ R_* \leq r \leq 0.960\ R_*$, for each of the 
solutions listed in Table 1.  It is apparent that rings with smaller values
of $\beta_0$ and $a_0$ generally traverse the radiative envelope in less time
than do rings that are less strongly magnetized and have larger cross-sectional
radii (see also Moss 1989).  Based on the discussion of the preceding subsection, 
rings having smaller values of $\beta_0$ and $a_0$ are initially more buoyant, and 
are more likely to remain so because of the comparatively shorter time it takes to 
heat them by inward diffusion of radiation.  With the exception of the rings
corresponding to solutions 7, 9, and 10, the time needed to arrive at the
stellar surface is less than the estimated main sequence lifetime of the
9 $M_\odot$ model, $t_{MS} \approx 10^{10}\ (M_*/M_\odot)(L_\odot/L_*)
\approx 2.4 \times 10^7$ years.  However, note that even the more slowly rising
tubes of solutions 7, 9, and 10 attain radii in the range (0.92-0.95) $R_*$
within a time $< t_{MS}$.  In fact, all of the solutions shown in Figure 9
spend more time crossing the outer 10\% of the stellar radius than they do
traveling from $r = r_0$ to $r = 0.9\ R_*$.  From equation (27), it can be seen 
that since $T_e$ decreases and $\kappa_e$ increases as $r$ approaches $R_*$ in the
outer envelope, $u_r$ becomes small and the rise time increases accordingly.

Examination of Table 1 and Figure 9 also reveals that the omission of virtual
inertia and aerodynamic drag from the equation of motion has little impact on ring
dynamics and rise times (see solutions 1, 3, and 4).  This is because the flux
rings considered herein do not experience significant, impulsive accelerations,
and never achieve large rise speeds for prolonged periods of time.  Instead, 
most of the ascent takes place at the slow, quasi-steady rate given by equation 
(24) (or the approximation [25]).  This behavior accounts for another notable
characteristic of the results shown in Figure 9, namely, that the rise time
profiles of several solutions are nearly identical to one another.  From equation
(26), it is readily seen that the dependence of the rise speed on the intrinsic
properties of a given flux ring is $u_r \propto (a^2\ \beta)^{-1}$.  Consideration
of the entries in Table 1 indicates that, as expected, solutions having a common
profile of $t_r(r)$ in Figure 9 are characterized by the same value of the parameter 
combination $(a_0^2\ \beta_0)^{-1}$.

Finally, we note that with the preceding developments, it is possible to derive
an approximate expression for the rise time $t_r(r)$.  To do this, we first use
the equation of hydrostatic equilibrium in the form ${\rm d} \rho_e / {\rm d} r
= - (g\ \rho_e) /c_e^2$, where $c_e$ is the external sound speed, to convert the 
integration variable in equation (25) from $r$ to $\rho_e$; this procedure yields
\begin{equation}
t_r (r) = - \int \limits^{\rho_e}_{\rho_{e0}} {\rm d} \rho_e\ 
\left( {c_e^2 \over \rho_e\ g\ u_r} \right)
\approx - \left( {c_{e0}^2 \over \rho_{e0}\ g_0\ u_{r0}} \right)
\int \limits^{\rho_e}_{\rho_{e0}} {\rm d} \rho_e\
\left( {\rho_e \over \rho_ {e0}} \right)^{(\alpha -2)}
\left( {T_{e0} \over T_e} \right)^3
\left( {\kappa_e \over \kappa_{e0}} \right). 
\end{equation}
To obtain the second, approximate equality in equation (26), we have replaced $u_r$
by the scaling relation (27) and used the fact that $c_e^2 = c_{e0}^2\ 
(\rho_e / \rho_{e0})^{(\alpha -1 )}$ where $c_{e0}^2 = (\alpha\ p_{e0} / \rho_{e0})$.
The integrand in equation (26) can be simplified by assuming that the opacity
follows Kramer's law, $\kappa_e \propto \rho_e  T_e^{-3.5}$.  This assumption,
together with the polytrope relation $T_e \propto \rho_e^{(\alpha-1)}$ (see Appendix B),
allows us to express the integrand as a function of $\rho_e$ only; performing the
integration we obtain
\begin{equation}
t_r(r) = \left( {2 \over 11\ \alpha -13}\right)\  
\left( {c_{e0}^2 \over g_0\ u_{r0}} \right)\ 
\left[ \left( {\rho_e \over \rho_{e0}} \right)
^{-{(11 \alpha -13) \over 2}} - 1 \right]. 
\end{equation}
If $u_{r0}$ is determined by evaluating equation (26) at $r_0$, the rise
time estimate (27) becomes
\begin{equation}
t_r(r) = 3.0835 \times 10^3\ 
\left[ \left( {a_0 \over h_{e0}} \right)^2\ \beta_0 \right]\ 
\left[ \left( {\rho_e \over \rho_{e0}} \right)^{0.9349} - 1 \right]\ 
{\rm years}, 
\end{equation}
where the interior model of Appendix B has been used to fix the values of all
physical quantities at the core-envelope interface.  As an example, the \
dashed line in Figure 9 represents the rise time profile obtained from 
equation (30) for $[(a_0/h_{e0})^2\ 
\beta_0] = 10^{-2}$

\section{Conclusions and Discussion}

The foregoing examination of the physics of magnetic flux rings in the radiative
envelope of a uniformly rotating hot star has revealed a range of possible
dynamical behaviors, and has enabled us to estimate the efficiency of buoyancy
as a means of transporting dynamo-generated fields from the core to the surface.
Rings located in the equatorial plane can attain a stationary equilibrium in which
$u_r = 0,\ u_\phi \neq 0$, and the combined Coriolis and tension forces balance
buoyancy.  This state is stable against infinitesimal displacements of
the ring in the radial direction, but unstable when the perturbations are
meridionally directed, causing the ring position to shift from latitude $0^\circ$.
Apart from transient episodes of oscillatory behavior at the outset of their
motion, rings released from latitudes other than $0^\circ$ move to larger radii
along paths that are very nearly parallel to the stellar rotation axis.  During
most of the ascent, the rate of rise of such a ring is controlled by the rate
at which it is radiatively heated, with a near balance prevailing between the 
accelerations produced by the Coriolis and magnetic tension forces in the $r$- 
and $\theta$-directions. 

Strongly magnetized rings with smaller cross-sectional radii experience larger
initial buoyant accelerations and have shorter radiative heating time scales.
As a result, they attain higher rise speeds and require less time to traverse
the envelope than do rings with weaker fields and/or larger cross-sectional
radii (see also Moss 1989).  A majority of the ring models enumerated in Table 1 
reach the stellar surface in a time that is less than the main sequence lifetime 
of the 9 $M_\odot$ star, the only exceptions being solutions with large values
of $\beta_0$ and/or $a_0$.  If the field strengths and flux tube sizes
considered herein are realistic, this result suggests that the field
produced by a core dynamo could manifest itself in the surface layers of a
hot star at a relatively early stage of main sequence evolution.  In light
of the discussion of \S3, the strength of the field in a buoyant flux
ring that reaches the photosphere of the 9 $M_\odot$ stellar model is
estimated to be $B_* \approx B_0\ (\rho_{e*} / \rho_{e0}) \approx 9.44\ 
(\beta_0 / 10^4)^{-1/2}$ G.

The preceding conclusions are drawn from results obtained using a simple
model for a thin, isolated, untwisted flux ring immersed in a non-evolving,
rigidly rotating stellar interior.  Given the exploratory character of this
investigation, a more detailed treatment of flux tube structure and dynamics
is neither warranted nor practical.  However, several potentially important effects
that have been neglected in the present analysis can be included within the context
of the basic model of \S2.  Among these are large-scale internal circulatory
flows, the development (over time) of gradients in composition and angular
velocity, the influence of radiation pressure on tubes inside more massive
stars, mass loss, and the consequences of magnetic interference with radiative 
energy transport (e.g., thermal shadows and small-scale flows; Parker 1984). 

Of the effects noted above,
rotationally driven, meridional circulation may have the most significant impact
on the results described in \S3.  As noted in that section, each of the ring
models listed in Table 1 spends considerably more time traveling through the thin
shell $0.90\ R_* \leq r \leq 0.96\ R_*$ than in ascending from the starting 
radius to $r = 0.90\ R_*$ (see Figure 9).  In solution 1, for example, the
former displacement requires approximately $1.29 \times 10^5$ years, while the 
latter occurs in only $1.38 \times 10^3$ years, a difference in travel time of
nearly a factor of 100.  We point out that the inclusion of meridional
circulation may decrease the time needed to transport magnetic flux through
the layers just below the photosphere.  Estimates of the circulation speed
inside upper main sequence stars suggest that while such flows proceed quite
slowly at great depth, they can become fast in the lower-density region near the 
surface (see, e.g., Kippenhahn \& Weigert 1994).  Hence, flux tube advection
by a circulatory flow may act to supplement buoyant transport in the outermost
portion of the stellar interior, thereby shortening the time between the production
of fields in the core and their emergence in the atmosphere.  The combined
effects of buoyancy and advection will be investigated in the next paper of
this series.

\acknowledgments 

We are grateful to Yuhong Fan, M. Maheswaran and  Paul Charbonneau for
helpful conversations, and to Steven Jackson for providing the
interior model used in the calculations, and to Wendy Mukluk for
the rising flux tube figure. We thank the referee for a useful question
regarding tube heating and for suggesting additional references.
JPC is also grateful  for a visiting scientist position at HAO.

\clearpage

\appendix

\section{ Derivation of the Ring Energy Equation}

To derive the energy equation for a thin, axisymmetric magnetic flux ring,
we begin with the first law of thermodynamics, written in the form,
\begin{equation}
{ {\rm d} {\cal E} \over {\rm d} t}
+ p\ { {\rm d} \over {\rm d} t} \left( {1 \over \rho} \right)
= { {\rm d} {\cal Q} \over {\rm d} t}, 
\end{equation}
where ${\cal E} = (p / \rho) /(\gamma -1)$ is the specific internal energy 
of the gas, and $({\rm d} {\cal Q} / {\rm d} t)$ is the rate at which heat
is input, per unit mass of ring material.  Let $F_e$ be the radiative flux
at the surface of the ring; in the case of interest to us here, $T < T_e$, 
and this flux is directed into the ring from the external medium.  We
assume that the ring is sufficiently opaque that all of the flux incident
upon it is thermalized within it. For a ring of mass $M$, the specific
radiative heating rate is then 
\begin{equation}
{ {\rm d} {\cal Q} \over {\rm d} t} =
{A\ F_e \over M} = {2\ F_e \over \rho\ a}, 
\end{equation}
where $A = 2 \pi a \cdot 2 \pi r\ {\rm sin}\ \theta$ is the surface area
of the ring.  If ${\bf \nabla} T_e$ is the local temperature gradient in the
surrounding stellar material and ${\bf n}$ is a unit vector normal to the ring 
surface, the diffusive flux of radiation into the ring is approximately
\begin{equation}
F_e = -{16 \over 3}\ {\sigma\ T_e^3 \over \kappa_e\ \rho_e}\ 
({\bf n} \cdot {\bf \nabla} T_e)
\approx {16 \over 3}\ {\sigma\ T_e^3 \over \kappa_e\ \rho_e}\ 
{\Delta T \over a}, 
\end{equation}
where $\Delta T = (T_e - T)$, and we have estimated $| \nabla T_e |
\approx (\Delta T /a)$. 

Explicit calculation of the derivatives appearing on the left-hand
side of equation (A1) leads to 
\begin{equation}
{ {\rm d} {\cal E} \over {\rm d} t}
+ p\ { {\rm d} \over {\rm d} t} \left( {1 \over \rho} \right)
= { p \over (\gamma - 1) \rho}\ { {\rm d} S \over {\rm d} t},
\end{equation}
where the quantity
\begin{equation}
S = {\rm ln} \left[ \left( {p \over p_0} \right)
\left( {\rho_0 \over \rho} \right)^\gamma \right],
\end{equation}
has been referenced to the initial thermodynamic conditions in the
ring.  Substitution of the results given in equations (A2), (A3),
and (A4) into equation (A1) yields the ring energy equation in 
the form used herein,
\begin{equation}
{ {\rm d} S \over {\rm d} t} = {32 \over 3}\ 
{ (\gamma - 1)\ \sigma\ T_e^4 \over \kappa_e\ \rho_e\ p\ a^2}\ 
{ \Delta T \over T_e} \equiv q. 
\end{equation}
The time scale $t_H$ for radiative heating of the ring can be expressed
in terms of the rate $q$ as $t_H \equiv [{\cal E} / ({\rm d} {\cal Q} /
{\rm d} t)] = q^{-1}$.

\section{\bf{ Approximate Stellar Structure}}

The distributions of $p_e$, $\rho_e$, and $T_e$ used in the
calculations described in \S3 are derived from a model for a
spherical, non-rotating star with mass $M_* = 9\ M_\odot$ and
radius $R_* = 3.678\ R_\odot$.  The star is chemically homogeneous
with composition $X=0.7131,\ Z=0.0169$, and contains a central,
convective core of mass $M_c = 0.308\ M_*$ and radius $R_c = 0.232\ 
R_*$.  The ambient physical conditions at the initial radial position
$r_0\ (= R_c)$ of the flux ring are $p_{e0} = 1.3250 \times 10^{16}$
dyne cm$^{-2}$, $\rho_{e0} = 4.7239$ g cm$^{-3}$, and $T_{e0} =
1.9976 \times 10^7$ K.  In order to enhance the precision of
computations involving ring properties that differ from those of
the external medium by very small amounts, it is helpful to use
an analytic approximation to the background stellar model.  We
discuss the derivation of one such representation in this appendix.

Unlike the convective envelope of the Sun, the radiative envelope of
the adopted stellar model contains a significant fraction of the 
mass of the star, $1 - (M_c /M_*) \approx 0.70$.
Consequently, the gravitational acceleration $g(r)$ in the inner
portion of the envelope does not vary simply as $r^{-2}$, but instead 
depends on the detailed distribution of mass.  We approximate $g(r)$
within the region between the convective core and the stellar surface
by using the following piecewise continuous fit to the computed interior
model:
\begin{equation}
\begin{array}{cccc}
 g(z) & = & g_1\ {\rm exp} \left[\dsp - \left(
{z - \xi_1 \over  \Delta_1} \right)^2  \right],& 
z_0 \leq z \leq z_1, \\[.5cm]
      & = & g_2\ {\rm exp} \left[\dsp - \left(
{z - \xi_2 \over \Delta_2} \right) \right],&
z_1 \leq z \leq z_2, \\[.5cm]
      & = & {{\dsp g_3} \over {\dsp z^2}},&
z_2 \leq z \leq 1, \\ 
\end{array}
\end{equation}
where $z \equiv r/R_*$ and
$$g_1 = 1.06 \times 10^5\ {\rm cm}\ {\rm s}^{-2},\ 
g_2 = 8.90 \times 10^4\ {\rm cm}\ {\rm s}^{-2},\ 
g_3 = 1.80 \times 10^4\ {\rm cm}\ {\rm s}^{-2},$$
$$\xi_1 = 0.27,\ \Delta_1 = 0.31,\ \xi_2 = 0.40,\ \Delta_2=0.34,$$
$$z_0 = 0.232,\ z_1 = 0.422,\ z_2 = 0.641.$$
We assume that $p_e$ and $\rho_e$ satisfy a polytrope relation
of the form $p_e \propto \rho_e^\alpha$ where the index $\alpha$
has a constant value.  This assumption, together with the
approximation for $g$ given in equation ($B1$), enables us to integrate
the equation of hydrostatic equilibrium in order to obtain the
structural properties of the stellar radiative interior.  The
results of this procedure are:
\begin{equation}
{\rho_e \over \rho_{e0}} = \left\{ 1 - \left( {\alpha -1 \over \alpha}
\right) \left( {\rho_{e0}\ g_1\ R_* \over p_{e0}} \right)
\left( {\sqrt{ \pi }\ \Delta_1 \over 2} \right)
\left[ \Phi \left( {z - \xi_1 \over \Delta_1} \right) -
\Phi \left( {z_0 - \xi_1 \over \Delta_1} \right)
\right] \right\}^{{1 \over (\alpha - 1)}}, 
\end{equation}
in $z_0 \leq z \leq z_1$, where $\Phi(x) = (2 / \sqrt{\pi})
\int \limits_0 ^x {\rm d}t\ e^{\dsp -t^2}$ is the error function;
\begin{equation}
{\rho_e \over \rho_{e1}} = \left\{ 1 - \left( {\alpha - 1 \over
\alpha} \right) \left( {\rho_{e1}\ g_2\ R_* \over p_{e1}} \right)
\left[ {\rm exp} \left( {\xi_2 -z_1 \over \Delta_2} \right) -
{\rm exp} \left( {\xi_2 - z \over \Delta_2} \right)
\right] \right\}^{{1 \over (\alpha - 1)}}, 
\end{equation}
in $z_1 \leq z \leq z_2$, where $\rho_{e1} = \rho_e (z_1)$ and
$p_{e1} = p_e(z_1)$; and,
\begin{equation}
{\rho_e \over \rho_{e2}} = \left[ 1 - \left( {\alpha - 1 \over
\alpha} \right) \left( {\rho_{e2}\ g_3\ R_* \over p_{e2}} \right)
\left( {1 \over z_2} - {1 \over z} \right) \right]
^{{1 \over (\alpha -1)}}, 
\end{equation}
in $z_2 \leq z \leq 1$, where $\rho_{e2} = \rho_e (z_2)$ and
$p_{e2} = p_e (z_2)$.  By virtue of the polytrope relation,
the pressure and temperature in each of the three intervals are
given by
\begin{equation}
{p_e \over p_{ei}} = \left( {\rho_e \over \rho_{ei}} \right)
^\alpha,\ {T_e \over T_{ei}} = \left( {\rho_e \over \rho_{ei}}
\right)^{\alpha - 1}, 
\end{equation}
where $i = 0,\ 1,\ 2$.

A polytrope with $\alpha = 1.3518$ yields a reasonable fit to the
properties of the computed stellar interior model.  Note that
a polytropic representation of this kind is not
completely consistent with the detailed model, a consequence of the
fact that the simple assumed relation between $p_e$ and $\rho_e$
provides only an average description of the entire stellar envelope.
A more accurate fit can be obtained if the index $\alpha$ is allowed
to have a distinct value in each of the three intervals delineated 
above (e.g., $\alpha_1 = 1.400$, $\alpha_2 = 1.330$, and
$\alpha_3 = 1.277$).  However, the single index description is
adequate for the purpose of the present computations.

\clearpage

\begin{figure}
\epsscale{0.7}
\plotone{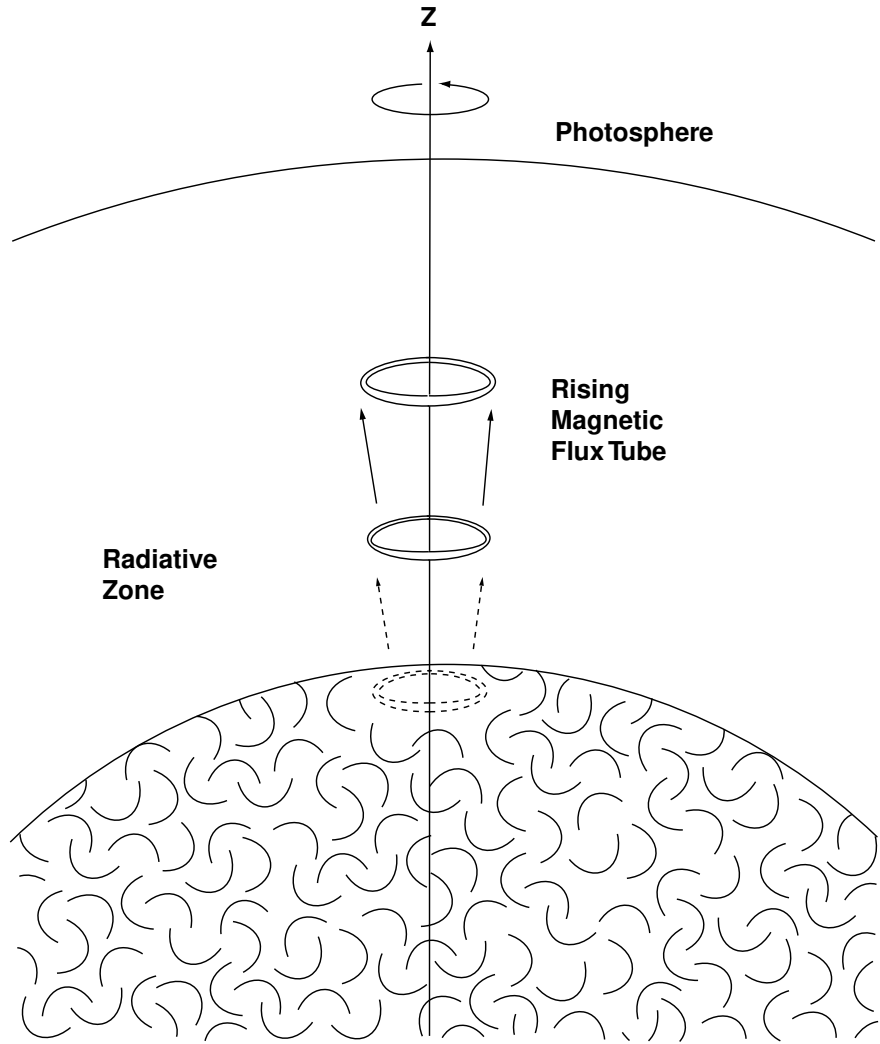}
%\figcaption[fig1] 
\caption[]
{An illustration of a flux tube rising through the radiative envelope
of a massive star. The tube is assumed to be a geometrically 
thin circular ring with its symmetry axis along the rotational
axis of the star. The field in the toroidal tube is assumed to 
be generated by a dynamo process acting at the interface of the 
convective core and  radiative envelope zones. The
tube is assumed to maintain the enclosed field and the initial 
mass of the tube.}
{ \label{risingtube} } 
\end{figure}

\clearpage

\begin{figure}
%\epsscale{1.0}
%\scalebox{0.65}[0.65]{\rotatebox{+90}{\includegraphics{f2.eps}}}
\centering\epsfig{file=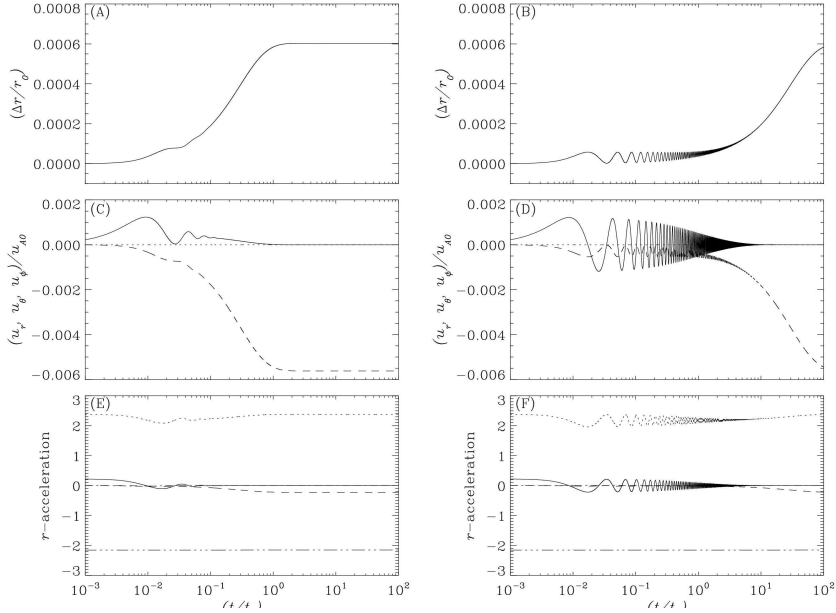,width=0.72\linewidth,angle=+90}
%\plotone{f2.eps}
%\figcaption[fig2] 
\caption[]
{The time evolution of selected properties of magnetic
flux rings with $\beta_0 = 10^4$ located in the stellar equatorial plane 
($\theta = \pi /2,\ \lambda = 0^\circ$).  The panels on the left-hand 
side of the Figure contain
results for a ring with initial cross-sectional radius $a_0 = 10^{-4}\ 
h_{e0}$, while those on the right-hand side pertain to a ring with $a_0
= 10^{-3}\ h_{e0}$.  Panels (A) and (B) show the normalized radial
displacement $(\Delta r /r_0)$ of the ring from its initial position $r_0$.
Panels (C) and (D) show the velocity components $u_r$ (solid line),
$u_\theta$ (dotted line), and $u_\phi$ (dashed line) in units of the
initial Alfv\'en speed $u_{A0}$ in the ring.  Panels (E) and (F) show
the radial accelerations (in units of $u_{A0}^2 / R_*$) produced by the 
buoyant (dotted line), inertial/Coriolis (dashed line), drag 
(dashed-dotted line), and magnetic tension (dashed-double dotted line)
forces.  The net radial acceleration is represented by the solid
line.}
{\label{rvelaccequator} } 
\end{figure}

\clearpage

\begin{figure}
%\epsscale{0.75}
%\scalebox{0.65}[0.65]{\rotatebox{+90}{\includegraphics{f3.eps}}}
\centering\epsfig{file=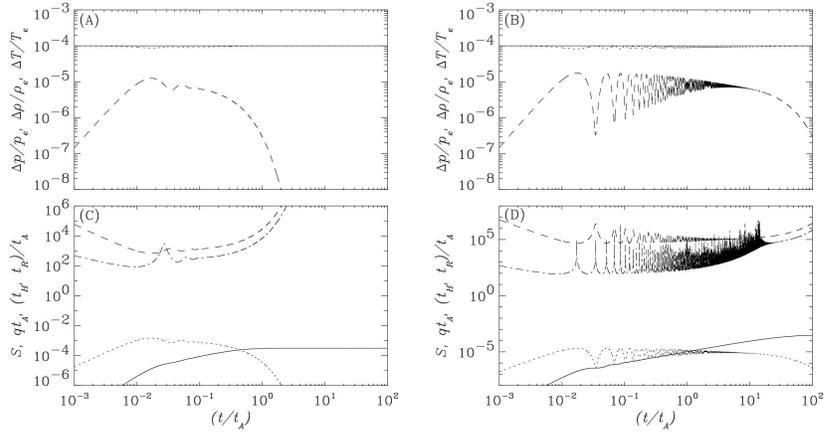,width=.72\linewidth,angle=+90}
%\plotone{f3.eps}
%\figcaption[fig3] 
\caption[]
{The thermodynamic evolution of the magnetic flux rings
depicted in Figure 2.  On the left-hand side of the Figure, $\beta_0 =
10^4$, $a_0 = 10^{-4}\ h_{e0}$, while on the right-hand side, $\beta_0 =
10^4$, $a_0 = 10^{-3}\ h_{e0}$.  Panels (A) and (B) show the normalized
pressure (solid line), density (dotted line), and temperature (dashed
line) deficits ($\Delta p$, $\Delta \rho$, and $\Delta T$, respectively)
of the ring material relative to the external medium.  Panels (C) and (D)
show the entropy-like quantity $S$ (solid line), the radiative heating
rate $q$ (dotted line), the heating time-scale $t_H$ (dashed line), and
the local rise time $t_R$ (dashed-dotted line).}
{ \label{thermoequator} } 
\end{figure}

\clearpage

\begin{figure}
%\epsscale{0.75}
%\scalebox{0.65}[0.65]{\rotatebox{+90}{\includegraphics{f4.eps}}}
\centering\epsfig{file=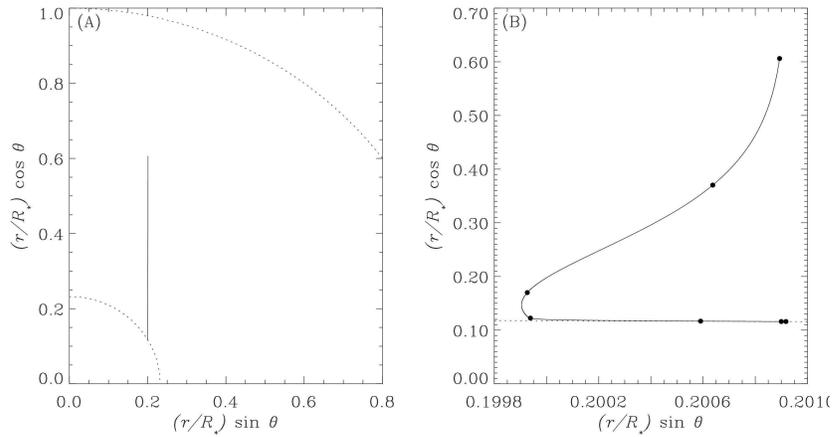,width=.72\linewidth,angle=+90}
%\plotone{f4.eps}
\caption[]
{The trajectory of a buoyant magnetic flux ring in the
radiative envelope of a 9 $M_\odot$ star, as described in \S3.2 of the
text.  Panel (A) shows the path followed by a ring with $\beta_0 = 10^4$
and $a_0 = 10^{-4}\ h_{e0}$, starting from a position at latitude
$\lambda_0 = 30^\circ$ ($\theta_0 = \pi /3$) on the core-envelope interface. 
The inner and
outer dotted lines denote the core and photospheric radii, respectively,
and the duration of the time interval between the first and last points on 
the path is $10^3\ t_A$.  The same trajectory is also shown in panel
(B), but with an expanded horizontal scale to accentuate changes in the distance
of the ring from the rotation axis of the star.  Beginning in the lower
right-hand portion of the panel, the dots along the path correspond to
the times $(t / t_A) = 10^{-3},\ 10^{-2},\ 10^{-1},\ 1,\ 10,\ 10^2$,
and $10^3$.}
{ \label{traject4-4} } 
\end{figure}

\clearpage

\begin{figure}
%\epsscale{0.75}
%\scalebox{0.65}[0.65]{\rotatebox{+90}{\includegraphics{f5.eps}}}
\centering\epsfig{file=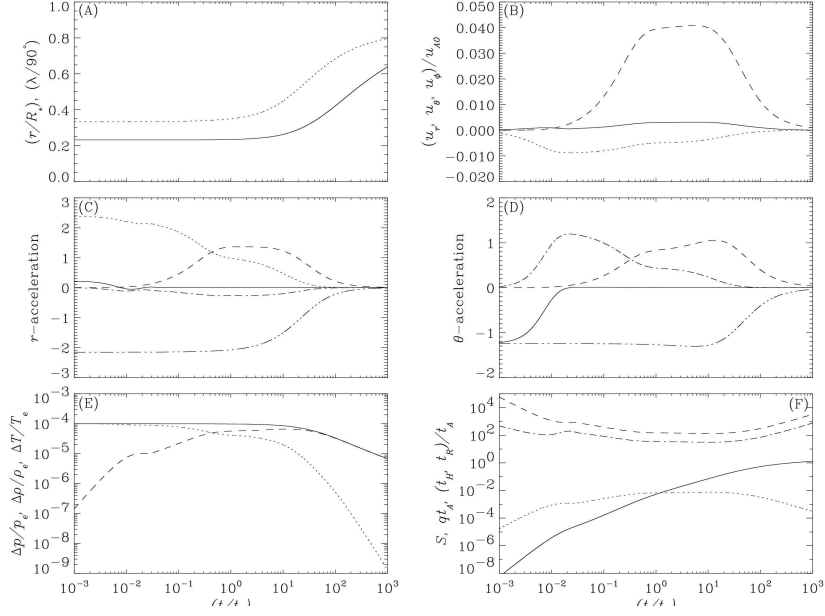,width=.72\linewidth,angle=+90}
%\plotone{f5.eps}
%\figcaption[fig5] 
\caption[]
{The dynamical and thermodynamical evolution of a magnetic
flux ring with $\beta_0 = 10^4$, $a_0 = 10^{-4}\ h_{e0}$, and $\lambda_0
= 30^\circ$.  All quantities are shown as functions of time, measured in
units of the Alfv\'en time $t_A$.  Panel (A) shows the normalized radial coordinate
$(r/R_*)$ (solid line) and latitude $(\lambda / 90^\circ)$ (dotted line)
of the ring.  Panel (B) shows the velocity components of the toroidal flux
tube, using the same format as in Figure 2.  Panels (C) and (D) show,
respectively, the accelerations produced by the $r-$ and $\theta-$components
of the forces acting on the ring.  Individual forces are identified
according to the line types used in the acceleration panels of Figure 2.
Panel (E) shows the contrasts in pressure, density and temperature
between the tube and its surroundings, while panel (F) shows $S$ and
the radiative heating rate, together with the local heating and rise
times.  The formats adopted for these panels are identical to those
used in Figure 3.} 
{ \label{rvelacc30degs} } 
\end{figure}

\clearpage

\begin{figure}
%\epsscale{0.75}
%\scalebox{0.65}[0.65]{\rotatebox{+90}{\includegraphics{f6.eps}}}
\centering\epsfig{file=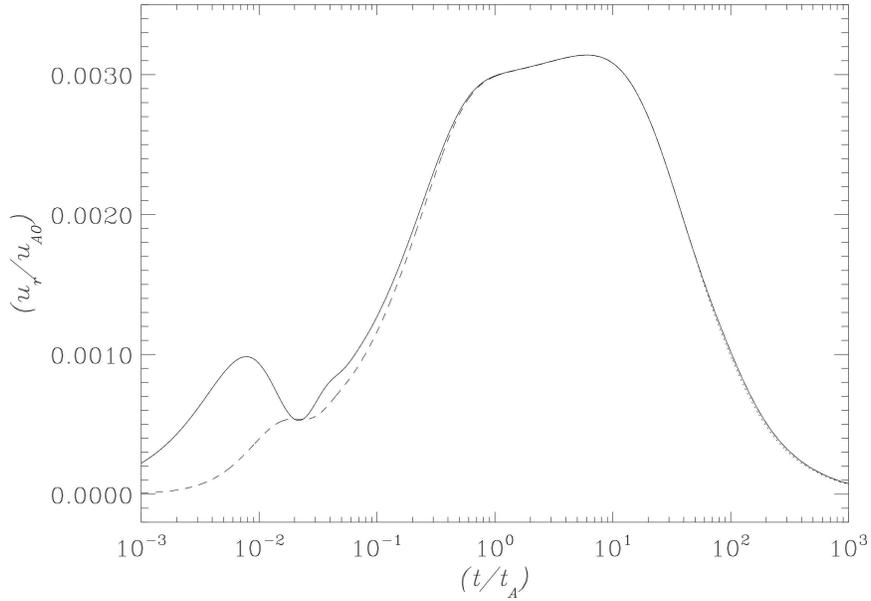,width=.72\linewidth,angle=+90}
%\plotone{f6.eps}
%\figcaption[fig6] 
\caption[]
{The time evolution of the radial velocity component $u_r$ for
a magnetic flux ring with $\beta_0 =10^4$, $a_0 = 10^{-4}\ h_{e0}$ and
$\lambda_0 = 30^\circ$.  The solid line corresponds to the radial rise speed 
derived from numerical integration of the equation of motion (5), while the dashed 
line represents the approximate rise speed given by equation (24).  The dotted line
for $t \geq 50\ t_A$ represents the rise speed given by equation (27), with the
constant of proportionality evaluated using the properties of the numerical
solution at $t = 50\ t_A$.}
{ \label{uvelvsanalyt} } 
\end{figure}

\clearpage

\begin{figure}
%\epsscale{0.75}
%\scalebox{0.65}[0.65]{\rotatebox{+90}{\includegraphics{f7.eps}}}
\centering\epsfig{file=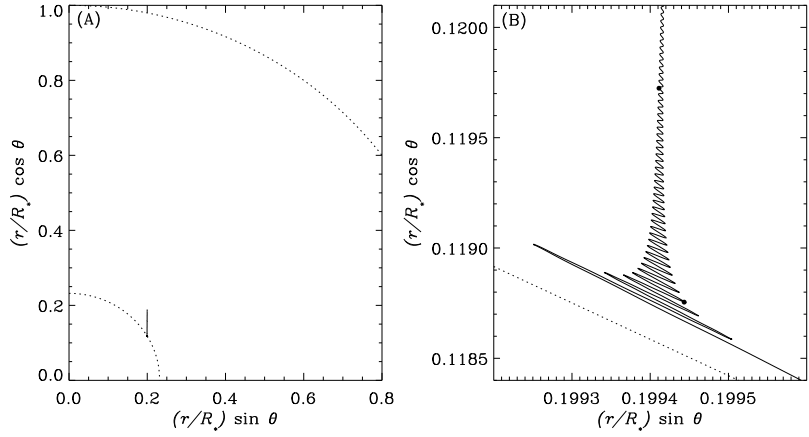,width=.72\linewidth,angle=+90}
%\plotone{f7.eps}
%\figcaption[fig7] 
\caption[]
{The trajectory of a magnetic flux ring with $\beta_0 = 10^4$,
$a_0 = 10^{-3}\ h_{e0}$, and $\lambda_0 = 30^\circ$, as in Figure 3.  The dots
on the path in the lower and upper portions of panel (B) indicate the positions
of the ring at times $t = t_A$ and $10\ t_A$, respectively.}
{ \label{traject4-3} } 
\end{figure}

\clearpage

\begin{figure}
%\epsscale{0.75}
%\scalebox{0.65}[0.65]{\rotatebox{+90}{\includegraphics{f8.eps}}}
\centering\epsfig{file=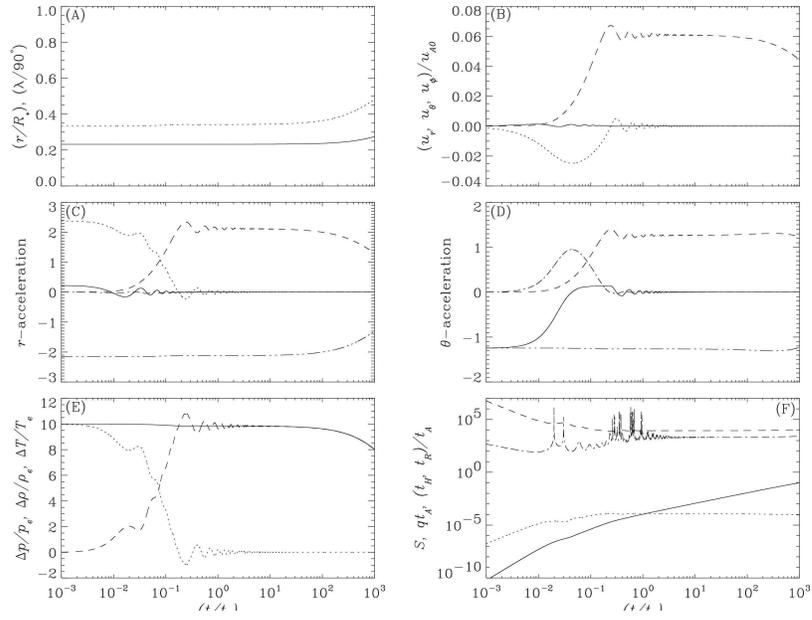,width=.72\linewidth,angle=+90}
%\plotone{f8.eps}
%\figcaption[fig8] 
\caption[]
{The dynamical and thermodynamical evolution of a magnetic
flux ring with $\beta_0 =10^4$, $a_0 = 10^{-3}\ h_{e0}$, and $\lambda_0
= 30^\circ$.  The format of the figure is the same as that of Figure 5,
with the exception that panel (E) depicts $10^5$ times the pressure,
density, and temperature contrasts, plotted on a linear scale to
accommodate the changes in sign that $\Delta \rho$ undergoes.}
{ \label{rvelacc4-3} } 
\end{figure}

\clearpage

\begin{figure}
%\epsscale{0.75}
%\scalebox{0.65}[0.65]{\rotatebox{+90}{\includegraphics{f9.eps}}}
\centering\epsfig{file=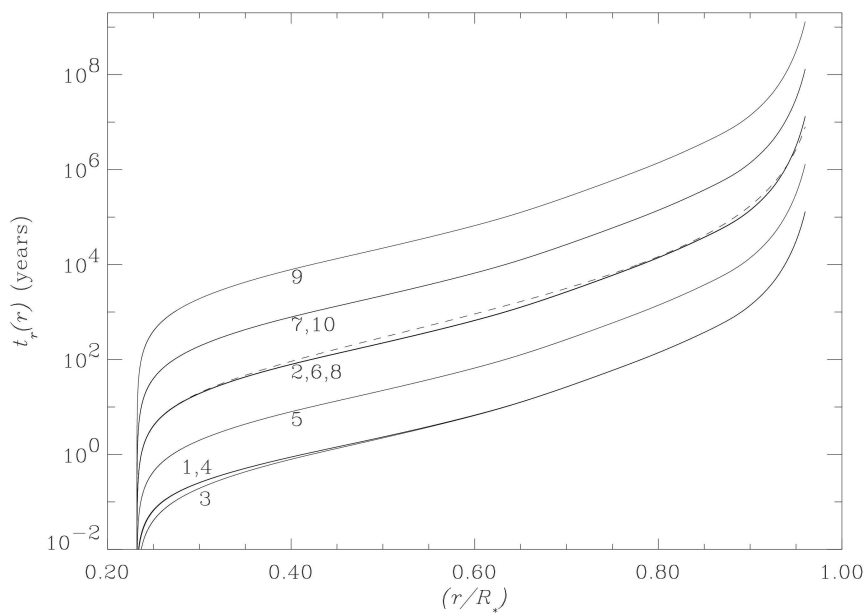,width=.72\linewidth,angle=+90}
%\plotone{f9.eps}
%\figcaption[fig9] 
\caption[]
{The time required for a magnetic flux ring to travel
from $r_0 = 0.232\ R_*$ to a given radius $r$ in the interval $r_0 \leq
r \leq 0.960\ R_*$.  The numbers labelling each curve correspond to the
models listed in Table 1.  The dashed line represents the rise time
estimate (30), evaluated for flux rings having $[ (a_0/h_{e0})^2\ 
\beta_0 ] = 10^{-2}$.}
{ \label{risetime} } 
\end{figure}

\end{document}